\documentclass[bibtex,11pt]{revtex4-1}
\usepackage{graphicx,amsmath,amsfonts,latexsym,bm,eso-pic}
\usepackage{bm}
\usepackage{amssymb}
\usepackage{color}
\usepackage{subfigure}

\def\nn{\nonumber}

\def\der#1#2{\frac{\partial #1}{\partial #2}}

\def \av#1{{\langle#1\rangle}}

\def \be{\begin{equation}}
\def \ee{\end{equation}}
\def \ba{\begin{array}}
\def \ea{\end{array}}
\def \bea{\begin{eqnarray}}
\def \eea{\end{eqnarray}}
\def \nn{\nonumber}
\def \l{\left}
\def \r{\right}
\def \rr{\right}
\def \half{{1\over 2}}

\def \H{{\cal{H}}}

\def \W{{\Omega}}

\def \e{{\epsilon}}

\def \a{{\alpha}}

\def \b{{\beta}}

\def \D{{\Delta}}
\def \d{{\delta}}
\def \w{{\omega}}

\def \f{\hat{\theta}}

\def \G{{\Gamma}}
\def \z{{\zeta}}

\def \on{\overline{n}}
\def \intt{\int\limits}

\def \av#1{{\langle#1\rangle}}

\def \summ{\sum\limits}
\def\intt{\int\limits}

\begin{document}

\title {Strong disorder renormalization group primer and the superfluid-insulator transition}

\author{Gil Refael}

\affiliation{Department of Physics, California Institute of Technology, Pasadena, California 91125, USA}

\author{Ehud Altman}

\affiliation{Department of Physics, University of California, Berkeley, California 94720, USA; and\\
Department of Condensed Matter Physics, Weizmann Institute of Science, Rehovot 76100, Israel}

\begin{abstract}
This brief review introduces the method and application of real-space renormalization group to strongly disordered quantum systems. The focus is on recent applications of the strong disorder renormalization group to the physics of disordered-boson systems and the superfluid-insulator transition in one dimension. The fact that there is also a well understood weak disorder theory for this problem allows to illustrate what aspects of the physics change at strong disorder. In particular the strong disorder RG analysis suggests that the transitions at weak disorder and strong disorder belong to distinct universality classes, but this question remains under debate and is not fully resolved to date. Further applications of the strong disorder renormalization group to higher-dimensional Bose systems and to bosons coupled to dissipation are also briefly reviewed.
\end{abstract}

 \maketitle

\section{Introduction}

Random systems can be broadly classified by the effect disorder has at large length scales. In many cases, the quenched randomness tends to be averaged out on long distances. Disorder then plays only a minor role in determining the universal behavior, and could be understood at a perturbative level. The more interesting alternative is that the disorder remains finite as the system is coarse grained. Finally, at the opposite extreme are systems in which the disorder grows without bound upon coarse graining. Such systems are said to be governed by infinite-randomness fixed points. 

The strong disorder renormalization group (SDRG) method, which is the focus of this review, provides a way to exploit the strong randomness in order to systematically compute universal aspects of the physics.   
The technique was originally developed by Dasgupta and Ma\cite{Ma1979,Dasgupta1980} to investigate the ground state and low energy behavior of the random Heisenberg spin chain. Later, the SDRG scheme was extended by Bhatt and Lee \cite{Bhatt1982} and formulated rigorously by Daniel Fisher\cite{Fisher1992,Fisher1994,Fisher1995}. Fisher showed in particular that the scheme gives asymptotically exact results for the low energy universal behavior of systems controlled by infinite randomness fixed points. For example the random spin-1/2 Heisenberg chain, flows toward a ground state decribed by an infinite randomness fixed point, the random singlet phase, for any amount of bare disorder\cite{Fisher1994}. Somewhat richer physics  is at play in the random transverse field Ising chain\cite{Fisher1992,Fisher1995}. This model exhibits a quantum phase transition between a magnetically ordered and a paramegnetic phase that is controlled by an infinite randomness fixed point. The analysis of the Ising model was extended to two dimensions by Motrunich et. al. \cite{FisherMotrunich}, who found a similar transition controlled by an infinite randomness fixed fixed point.  

The SDRG approach has since been applied and extended to address a host of problems spanning different fields. These include classical stochastic dynamics\cite{Fisher1998}, Dynamic response of spin chains at low temperatures\cite{Damle2000}, entanglement in random spin chains\cite{Refael2004} and most recently non-equilibrium quantum dynamics \cite{Vosk2013}. For a comprehensive account of the technique and its various applications, we refer the reader to Ref. \cite{Igloi2005}.

The goal of this brief review is twofold. First, we aim to provide a clear and compact introduction to the SDRG method and the physical considerations involved in its application. The second objective is to review recent progress in understanding the superfluid-to-insulator transition of interacting bosons in a random potential from the standpoint of SDRG.  

In contrast to other problems mentioned above, it is not immediately clear why the SDRG should be a suitable approach to characterize the superfluid-insulator transition. In fact an accepted theory of boson localization in one dimension is perturbative in the disorder\cite{Giamarchi1987,Giamarchi1988}. It predicts a critical point at which disorder is (dangerously) irrelevant. However, it has recently been argued that the superfluid may undergo a very different localization transition if the bare disorder is sufficiently strong\cite{Altman2004,Altman2010}. Moreover, as we shall discuss, the superfluid  phase itself may become anomalous due to the strong disorder. 
 
The rest of the review is structured as follows. In section \ref{sec:method} we review the technique using the spin-$1/2$ Heisenberg chain as an example to illustrate the scheme and how it is used to extract universal physical properties. In section \ref{sec:bosons} we turn to the application of SDRG to the superfluid insulator transition of bosons in a random one dimensional potential. We contrast the strong-disorder theory with the weak disorder analysis of Giamarchi and Schulz \cite{Giamarchi1987,Giamarchi1988}. In section \ref{sec:extensions} we discuss extensions of the SDRG analysis to bosonic systems with ohmic dissipation and to bosons in two dimensions.

\section{Brief review of the technique }\label{sec:method}

 In the standard application of SDRG to random systems the aim is to solve for the universal properties of the ground state and low energy excitations. As in any RG scheme, we focus on low energies by successively eliminating high energy modes, thereby generating a series of effective Hamiltonians  acting on the thinning Hilbert space. Within field theory, this program is often facilitated by the weakness of the non-linear coupling that mixes high and low frequencies. What is the guiding principle that allows to safely eliminate high energy modes of a random system without changing its low energy physics? 

The key  is the local separation of scales effected by the strong randomness. A grain of the system with atypically high energy near the cutoff scale $\W$, is likely to be surrounded by much weaker couplings.  The broader the disorder distribution
the more the chosen grain sticks out of its surrounding, thereby allowing a perturbative treatment of the coupling between the grain and its lower energy neighbors. Lets see how it works in a concrete example.

\subsection{Dasgupta-Ma decimation in the spin-1/2 Heisenberg chain}\label{sec:decimate}
We now explain the procedure through the simple example of a random spin-1/2 Heisenberg chain\cite{Ma1979,Dasgupta1980}. 
\be
H=\sum_{i}J_i {\bf S}_i\cdot {\bf S}_{i+1}
\ee
The motivation to study this model came from experiments done on
quasi-1d organic salts, mostly $Qn (TCNQ)_2$ \cite{Bulasevskii1972,Tippie81}. These salts have
chains of stacked double benzene rings, with each pair having one
excess spin-1/2. The measured
 susceptibility behaves as a power-law $\chi\sim T^{-\alpha}$
with $\alpha<1$ and varying from sample to sample. Recall that the Currie susceptibility of free spins is $\chi\sim
T^{-1}$.

Let us assume that the exchange coupling $J_i$ between the spins is widely distributed. Somewhere on a finite chain there is a bond $l$ having the largest exchange coupling which we denote by $J_l=\Omega$ (From now on we denote the upper energy cutoff of the chain by $\W$). Because the distribution is wide, the largest bond is much stronger than a typical bond and in particular it is likely to be much larger than the neighboring bonds, that is 
$J_{l\pm 1}\ll \Omega$. Therefore in our search for the ground state of the chain we can first diagonalize the strongest bond with $H_0=\W\,{\bf S}_l\cdot {\bf S}_{l+1}$ and treat the couplings to the rest of the chain as a perturbation. 

At zeroth order the degenerate low energy manifold of the chain consists of the spins $l,l+1$ frozen to a singlet state, whereas all other spins of the chain are free. The effective Hamiltonian acting within this subspace
is obtained through degenerate perturbation theory in all the other couplings. At first order we retrieve all the original nearest neighbor couplings along the chain, except the coupling of the strong pair to their left and right neighbors. A new coupling between the left and right neighbor spins is generated at second order of perturbation theory through virtual occupation of the excited states of the strong pair at energy $\W$:
\be
{\tilde H}_{l-1,l+2}={J_{l-1} J_{l+1}\over 2\W}{\bf S}_{l-1}\cdot{\bf S}_{l+2}.
\ee

We thereby eliminate the two strongest interacting
spins and reconnect the chain by generating an effective coupling between the spins to the left and right
of the decimated bond: 
\be
J^{\text{eff}}_{l-1,l+2}=\frac{J_{l-1}J_{l+1}}{2\Omega}\ll J_{l-1},J_{l+1}\ll\W
\label{Jeff}
\ee
 We expect this perturbation theory, controlled by the small parameter $J_{l\pm1}/\W$,  to work almost every time in the limit of strong disorder. Crucially, after the decimation step we
have exactly the same form of the Hamiltonian: a nearest-neighbor
Heisenberg model. 

The next stage in our analysis must be repeated application of the
decimation step. This leads to gradual renormalization of the distribution of the exchange constants $J_i$ upon decreasing the cutoff $\W$. If the disorder increases with repeated decimation, then we are safe. The strong
disorder assumption only becomes better and better. This is indeed the
case as we will show below.



\subsection{Qualitative ground state picture: the random singlet
  phase}

\begin{figure}\begin{center}
\includegraphics[width=12cm]{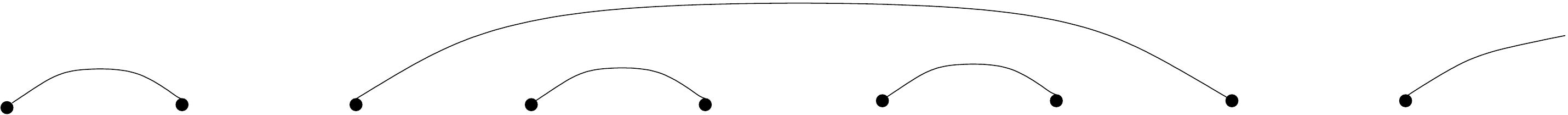}
\caption{The random singlet phase of a random Heisenberg model. Pairs of
strongly interacting sites form non-overlapping singlets in a random fashion. These
singlets mostly form between nearest neighbors, but also over an
arbitrarily large distance. The long range singlets induce strong
correlations between far away sites. \label{rsfig4}}
\end{center}\end{figure}

Before going into the formal derivation of the flow equations, let us sketch a qualitative picture of the ground state that may be inferred from the structure of the  RG decimation rules. A decimation of a bond essentially freezes two spins in a singlet state. At early stages of the RG, many singlets form between nearest neighbors. However, as more bonds are decimated, and the cutoff scale $\W$ is reduced, the largest couplings may occur on bonds generated at earlier stages of the RG between further
neighbors. Eventually as $\Omega$ is reduced far below its initial value singlets may form between  very far sites. The qualitative picture of the ground state generated in this process shown in 
Fig. \ref{rsfig4} looks like a random arrangements of non crossing singlet pairs occuring at all scales. Hence the name random-singlet phase.

From this simple picture we can infer important information about the nature of the correlations in the system. Suppose we are given a single realization of  a random Heisenberg  chain. What is the typical correlation $C_{ij}=\av{{\bf S}_i\cdot{\bf S}_j}$ we will measure between two far removed spins $i$ and $j$? Most likely these specific spins have not formed a singlet and therefore the correlation between them is very small - exponentially suppressed with distance (as we shall see later, with the square root of the distance):
\be
C_{ij}^{typical}\sim e^{-a \sqrt{|i-j|}}.
\ee 
These are called the typical correlations. 

What if rather than a single realization the measurement averages over an ensemble of chains. This would be the case, for example, if we could do neutron scattering on a bulk sample of the material $Qn (TCNQ)_2$, which contains a macroscopic number of chains.
The average over the many realizations could be dominated by rare instances in which the correlation $C_{ij}$ is atypically strong. Specifically, in the rare possibility that the two spins $i,j$ do happen to form a singlet, the correlation between them would be $-1$.
How rare is that really? If the two sites considered survive through
the decimation procedure until they become nearest neighbors, then they are very likely
to form a singlet. The probability of a site to survive to that stage, i.e., until $|i-j$ of its original nearest neighborson the side of the other site are removed, is $1/|i-j|$ (this is the density of survivng sites at that stage). Hence the probability of both sites $i$ and $j$ to survive to that stage is  $p_s\approx 1/|i-j|^2$.
We can now compute the average correlation to be
\be
\overline{C}_{ij}=(-1)p_s+e^{-a\sqrt{|i-j|}}(1-p_s)\approx -p_s\sim {1\over (i-j)^2},
\ee 

So despite the localized nature of the ground state, the average
correlations fall off only as a power law. This is a good example
of {\it Griffiths effects} \cite{Griffiths1969}, where the average correlations of a
random system are dominated by rare instances with anomalously strong
correlations. 

\subsection{Master equation for the flow of distribution functions}

Let us now show how detailed and precise information about the ground state and low-energy correlations is obtained, following Fisher \cite{Fisher1994}. The key step is to translate the Dasgupta-Ma decimation rules into a master equation describing how the repeated decimations renormalize the probability distribution of exchange couplings when they operate on an ensemble of Hamiltonians. 

Let us denote by $\rho_0(J)$ the distribution of $J$ in the physical system of interest. As we decimate more bonds while reducing the cutoff scale from $\W_0$ to $\W$ the distribution evolves to $\rho(J,\W)$. To derive the master equation that governs this evolution, it is much better to replace $J$ and $\G$ by the dimensionless scaling variable $\zeta_i=\ln ({\W/ J_i})$ and RG scale parameter $\G=\ln(\W_0/\W)$.  While $J$ was defined
on the changing interval $[0,\W]$, $\z$ is
always defined on the interval $[0,\infty]$. In particular $\z_i=0$ corresponds to the bond with largest exchange coupling $J=\W$. In these variables the Dasgupta-Ma decimation rule (\ref{Jeff}) takes the additive form:
\be
\z_{i-1,i+2}=\z_i+\z_{i+1}+\ln2
\label{DasMaLog}
\ee
Due to the strong disorder assumption, $\z_i =\ln (\W/J_i)$ is almost always much greater than $-\ln 2$, and it will therefore be safe to ignore the latter in Eq. (\ref{DasMaLog}).

Now we are ready to derive the master equation for the distribution of $\zeta$'s - which we denote $P_{\G}(\zeta)$. Integrating out the high energy shell $[\W-d\W,\W]$ consists of two stages: (i) Remove all the strong bonds with $0<\z<d\G$ and redefine the remaining $\zeta$'s according to the new cutoff, (ii) add the couplings generated through second-order perturbation theory across the decimated bonds.

Let's consider the contribution of stage (i) to the change of the distribution function. Having reduced the cutoff we need to redefine the  $\zeta_m$ on every remaining bond:
\be
\zeta_m\rightarrow \z_m-d\z=\ln
\frac{\Omega-d\Omega}{J_m}=\zeta_m-\frac{d\Omega}{\Omega}=\zeta_m-d\G.
\ee
So the entire distribution $P(\zeta)$ moves to the left. This can be
expressed mathematically:
\be
dP(\zeta)=\der{P(\zeta)}{\zeta}d\G.
\label{m1}
\ee

Now consider the contribution from stage (ii), which is adding the renormalized
bonds. The RG rule (\ref{DasMaLog}) prescribes how the distribution of the new bonds derives from that of the original constituent (left and right) bonds:
\be
P_{\text{new}}(\z)=\int_0^\infty d\z_l\int_0^\infty d\z_r P(\z_l)P(\z_r)\d(\z-\z_l-\z_r-\ln2)
\ee
The contribution of these bonds to the full distribution should be scaled by their fraction in the total population. Since the new bonds are produced only where we find a strong bond the probability of generating them is the probability to find a strong bond $\z\in[0,\G]$ that is  $P(0)d\G$. From this we get the  contribution of the newly generated bonds to the full distribution
$dP(\z)=d\G P(0)P_{\text{new}}(\z)$. 

Putting the two contributions together we obtain the master equation:

\be
\frac{dP(\zeta)}{d\G}=\der{P(\zeta)}{\zeta}+ P(0)\int_0^\infty d\zeta_{\ell}\int_0^\infty d\zeta_{r}
P(\zeta_{\ell})P(\zeta_{r})\delta(\zeta-\zeta_{\ell}-\zeta_r),
\label{me}
\ee
where we dropped the $\G$ subscript of $P(\zeta)$ and neglected the $\ln 2$ in the $\d$-function. 

There is a point we glossed over. We removed some probability by
getting rid of all the probability density at small $\zeta$, and we added
probability by adding all the new bonds. Do we need to adjust the
normalization of our distribution function? Integrating both sides of
Eq. (\ref{me}) reveals that the normalization is unchanged. For each
bond we lost, we added a renormalized bond.

\subsection{Solution of the flow equation}

The additive nature of the decimation rule  (\ref{DasMaLog}), embodied in the $\d$-function that appears in the master equation (\ref{me}) suggests a potential solution in the form of an exponential distribution $P_\G(\z)=f(\G)e^{-f(\G)\zeta}$. Indeed, plugging this ansatz into (\ref{me}) leads to an ordinary differential equation for $f(\G)$, $\partial_\G f=-f^2$, which is solved by $f(
\G)=1/\G$. Hence we obtain the self similar solution 
\be
P_\G(\zeta)={1\over \G}e^{-\z/\G}.
\ee
We see that the system flows to infinite randomness as the width of the 
distribution grows without limit.  
Converting back to the physical variables we obtain a power law distribution of effective exchange coupling
\be
\rho_\W(J)= {1\over\W\G} ({\W\over J})^{1-1/\G}
\label{rs-dist}
\ee
This approaches a non-normalizable distribution $\propto 1/J$ at the fixed point. Again a sign of infinite randomness.
 
The distribution (\ref{rs-dist}) is an attractor of the RG flow. Moreover, it turns out to be a global attractor.  The nearest neighbor Heisenberg chain flows to the same infinite randomness fixed point regardless of the initial distribution as long as the disorder is not correlated. Hence the results we will extract for the low energy physics are universal.

\subsection{Physical properties}

An important step in calculating physical properties is to establish a relation between energy scale $\W$ and length scale. A typical length scale is the distance between surviving spins at the scale $\W$.  
Lets compute the number of surviving spins on the original chain. Every time we decimate a strong bond, we remove two. Therefore,
upon changing the RG scale by $d\G$, the number changes by $dN=-2P_\G(0)N d\G$. Recall from the above solution that $P(0)=f(\G)=1/\G$; hence $N(\G)\sim N_0/\G^2$.
The average distance between surviving spins is then $L(\G)=l_0 N_0/N\sim \G^2$, or converting to energy units: $L(\W)\sim l_0\ln^2(\W_0/\W)$, where $l_0$ is the original lattice spacing.
This means that the {\it excitation energy} of singlets of length
$\ell$ is:
\be
\ln J_{\ell}\sim -\sqrt{\ell},
\ee
which is contrary to usual quantum-critical point scaling where $E\sim
1/\ell^{z}$. 
this type of scaling is called infinite-randomness scaling. 

We are now in position to compute the spin susceptibility at temperature $T$.
Lets run the RG decimation from the upper cutoff $\W_0>>T$ down to $\W=T$.
The decimated spins are essentially frozen into singlets and therefore do not contribute to the susceptibility. On the other hand, surviving spins at the scale $\W$ are typically coupled by bonds $J\ll T$, and therefore expected to behave as free spins. They contribute a Currie susceptibility
$
\chi(T)=n(T)/T
$, where $n(T)$ is the density of surviving spins on the chain at energy scale
$\W=T$.  From our previous calculation we have $n(T)=N(T)/N_0\sim n_0\ln^{-2} (\W_0/T)$. Therefore,
\be
\chi(T)=\frac{n_0}{T \ln^2 ({\Omega_0}/{T})}.
\ee
This is not quite $T^{-\alpha}$, but on a log scale it is indistinguishable from a power law.

\section{Random bosons: from infinite to finite randomness}\label{sec:bosons}

Let us now turn to the second aim of this review: the application of the SDRG method to
interacting bosons propagating in a random potential.

\subsection{Model}

Our analysis focuses on the quantum rotor Hamiltonian:
\be
\H=\sum_j U_j \left(\hat{n}_j-\overline{n}_j\r)^2 -\sum_j J_j
\cos\l(\hat\theta_{j+1}-\hat\theta_j\r), \label{model}
\ee
where on each site the phase $\hat\theta_i$ and charge $\hat n_i$ are conjugate variables, which obey $\l[\hat{n}_j,\,\f_k\rr]=-i\delta_{jk}$.  This model describes an effective Josephson-junction array (see Fig. \ref{fig:model}a) with random Josephson
couplings $J_i$ and charging energies $U_i$. In addition, there is a random
offset charge
${\bar n}_i$ in each grain, which is tantamount to a random gate voltage,
$V_j=2U_j\overline{n}_j$.
The integer part of $\on_i$ can be absorbed into the definition of $\hat n_i$ such that $\on_i$ is defined on the interval $(-\half,\half]$. In this review we focus on the case of generic disorder, where $\on$ can attain any value in this interval. The generic case, as well as the more restricted $\on$ disorder classes, were considered in Ref. \cite{Altman2010}. The different classes of disorder give rise to distinct insulating phases, but cause nearly unnoticeable differences in the behavior at the critical point.

Besides Josephson junction arrays there are several other important physical
realizations of the model (\ref{model}). In systems of ultra-cold trapped
atoms, disorder can be generated by optical speckle patterns\cite{Billy2008}, incommensurate lattice potential\cite{Lye2005}, or 
by corrugation in the wire that generates the trapping magnetic field in atom chips \cite{Kruger2007,Wang2004}. With increasing disorder, the atoms concentrate in small
puddles at minima
of the potential connected by random Josephson coupling
which depends
on the potential barrier between them. The Hamiltonian  (\ref{model}) can
be rigorously derived and the distributions of coupling constants computed
ab-initio \cite{Vosk2012}. Finally, disordered superconducting nano-wires
with no unpaired gapless electrons will also be generically described by Eq.
(\ref{model}).

More generally, the model (\ref{model}) is a good effective description for superfluid to insulator transition driven by phase fluctuations.
It can be thought of as a low energy effective theory for the phase degrees
of freedom, after  integrating out the gapped amplitude fluctuations. 

\subsection{The weak disorder limit\label{wdRG}}

Before describing the strong-disorder theory, let us briefly review the common understanding of the superfluid insulator transition in one dimension, taking the weak disorder viewpoint. The weak disorder theory was  formulated in a seminal paper by Giamarchi and Schulz  (GS) Ref. \cite{Giamarchi1987,Giamarchi1988}. 

The weak disorder expansion is natural to carry out when the action is written in terms of charge variables that live on the bonds rather than on the sites. Define
\be
\phi_j=\summ_{i\le j} n_i.
\ee
The current through the $i$'th bond is then
$\dot{\phi}_i$, and the charge on the $i$'th site is $\phi_i-\phi_{i-1}
$.
Therefore, the charging energy of the site is 
$E_{c}=U_i \l(\phi_i -\phi_{i-1}-\on_i\rr)^2$.

What about the Josephson energy? If we neglect phase slips, a Josephson junction is essentially an inductor with inductance $L_i=1/J_i$. Therefore the Josephson energy in this approximation is:
$
 E_J\approx L_i I^2/2={\dot\phi_i}^2{/ (2J_{i})}.
$
We are tempted to write the Lagrangian in this representation as
\be
\mathcal{L}_0=\sum_i \left[ \frac{1}{2 J_i} {\dot\phi_i}^2-U_i(\phi_i-\phi_{i-1}-{\bar n}_i)^2\right]
\label{GS-latt}
\ee
But this expression does not capture the periodicity of the Josephson energy with respect to the phase difference accross the junction. To account for the periodicity, we must allow for phase slips that change  $\theta\to\theta\pm2\pi$.  

Since the phase and charge are canonical conjugates, the translation of all phases to the left of the junction $i$ by $\pm2\pi$ is achieved by the operator
$\exp\left[\pm 2\pi i\sum_{j<i}n_j\right]=e^{\pm 2\pi i \phi_i}$.
Such terms should be included and assigned an action cost:
\be
\mathcal{L}_{ps}=- \summ_i \xi_i \cos\l(2\pi\phi_i\rr)
\ee
$\xi_i$ is the fugacity (or rate) of phase slips, and strongly
depends on both $J_i$ and $U_i,\,U_{i+1}$.  The total Lagrangian is given by $\mathcal{L}=\mathcal{L}_0+\mathcal{L}_{ps}$.

Within the weak disorder limit, the randomness in $1/J$ and $U$ that appears in $\mathcal{L}_0$ is perturbatively irrelevant. Furthermore,
the disorder in $\on_i$ can be absorbed by shifts to $\phi$. These shifts
then appear in the $\xi$ term, and produce a combination of  cosine and sine
terms. Rewriting the Lagrangian in the continuum limit, with this in mind,
gives
\be
{\mathcal{L}}\approx\frac{1}{2K}\int dx \left(\frac{1}{v} \dot{\phi}^2-v  \left(\nabla
\phi\right)^2\right)-\int{ dx}\,a^{-1}\left( \xi(x)  e^{2\pi i\phi}+\xi(x)^* e^{-2\pi i\phi}\right)
\label{GS-cont}
\ee
with $\xi(x)$ a complex number, $a$ is the original lattice constant, $v$ the sound velocity, and $K$ the Luttinger parameter (our definition of $K$ is the inverse of the $K$ appearing in Ref. \cite{Giamarchi1988}).  

The weak-disorder analysis proceeds by assuming a Gaussian distribution of the disorder with  $\overline{\xi(x)\xi(x')^*}=(\pi v)^2 a^{-1} D\delta(x-x')$. The disorder strength is parameterized by the dimensionless parameter 
 $D$. Next, momentum shell RG is
used to find the flow of $K$ and $D$. The result, quoted from \cite{Giamarchi1987}
is:
\be
\frac{dD}{dl}=D (3-2K^{}),\,\,\,\frac{dK}{dl}=-\half D K^2.
\ee
These scaling equations imply that the critical point is characterized by a {\em universal} value of the Luttinger parameter $K=3/2$ irrespective of the initial disorder strength. 

It is important to note, however, that obtaining the long-wavelength field theory (\ref{GS-cont}), with a single well defined Luttinger parameter $K$, relied on having weak disorder in $J$ and $U$ in the first place. We will see later what can go wrong with this mapping at strong disorder.



\subsection{Decimation steps for SDRG \label{rgsteps}}

Our main task now is to develop a strong-disorder RG scheme for the random-boson problem. As in the original application of real space RG to disordered systems, the strategy for finding the ground state and  low lying excitations is
to iteratively isolate and solve the strongest elements in the Hamiltonian.
 There are two types of elements in the Hamiltonian: onsite charging energies,
and bond Josephson couplings. Roughly, two sites connected by the strongest
bond will be converted to a
phase-coherent cluster. Similarly, in sites with strong charging
interactions we will eliminate all the on-site excited states. 

First, we must determine what we mean by strong elements of the Hamiltonian.
The charging interaction can be considered while ignoring Josephson couplings.
It is natural to pick the charging-energy gap separating the ground state
from excited charge states as the energy scale for decimation purposes. This
gap is given by:
\be
\Delta_i=U_i\l(1-2|\on_i|\r)
\label{cgap}
\ee
where $-1/2<\on<1/2$. At the same time, the energy scale, which
characterizes a bond is the Josephson coupling, $J_i$. A strong bond is
expected to bind two sites into a phase-coherent cluster. 

\begin{figure}[t]
 \centerline{\resizebox{0.7\linewidth}{!}{\includegraphics{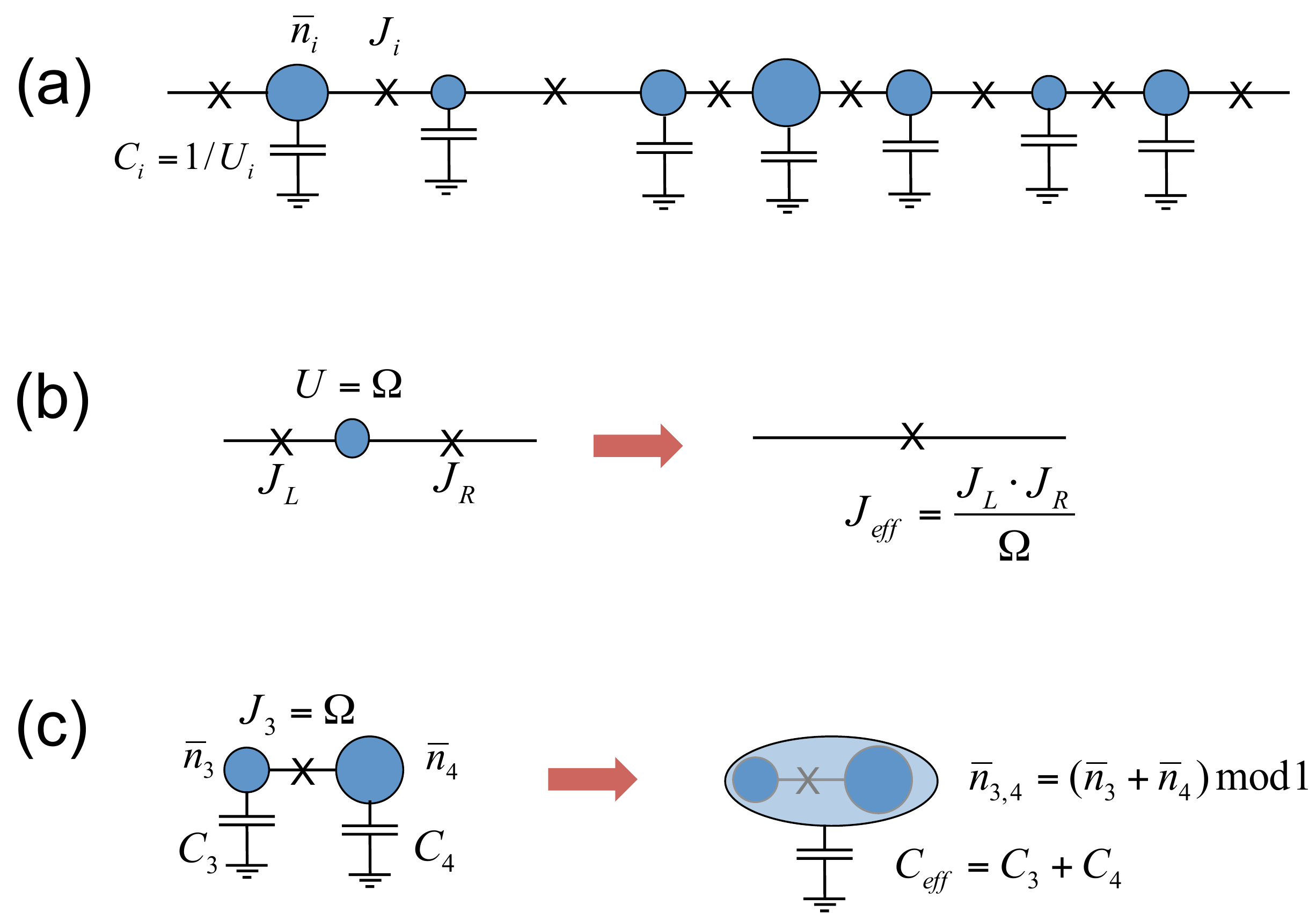}}}
 \caption{\em (a) The effective Josephson array model. (b) RG decimation of a large Josephson link. The connected islands are made into a single site with the sum of the two capacitances. The offset charges are added modulu 1.  (c) Decimation of a site with large charging energy. A new Josephson link is generated between the neighboring sites to the left and right of the decimated site. }
 \label{fig:model}
\end{figure}

At each step of the RG we eliminate the term responsible to the largest energy scale 
\be
\W=\max_i\{J_i,\,\Delta_i\}.
\label{ogap}
\ee
If the largest energy scale is a gap, $\Delta_i$, then the site $i$
freezes into the charge state with the lowest energy. Quantum fluctuations
induce an effective hopping between sites $i+1$ and $i-1$:
\be
J_{i-1,\,i+1}=\frac{J_{i-1}J_i}{\W (1+2|\on_i|)}.
\ee
This result, obtained by simple application of second order
degenerate perturbation theory, is illustrated in Fig. \ref{fig:model}b. The factor $1+2|{\bar n}|$ in the denominator varies between $1$ and $2$; it does not affect any universal features of the transition and we can safely set it to
unity. 

Alternatively, when the strongest coupling in the chain is the bond $J_i$,
a phase-coherent cluster forms, with a phase ${\tilde\theta_{i,i+1}}=(\hat\theta_i+\hat\theta_{i+1})/2$. Since charging energy is the inverse of
capacitance, and the capacitance is additive, the effective $U_{i,\,i+1}$
of the new cluster will be:
\be
\frac{1}{\tilde{U}_{i,i+1}}=\frac{1}{U_i}+\frac{1}{U_{i+1}}.
\label{urecurs}
\ee

The offset charge $\overline{n}$ in the cluster that forms in the bond
decimation step, is simply the sum of the two offset charges:
\be
\on_{i,\,i+1}=\on_i+\on_2.
\ee
This equality, however, is defined modulo adding or subtracting one,
so the the result always belong to the interval $\on_{i,\,i+1}\in(-1/2,1/2]$. This decimation step is illustrated in Fig. \ref{fig:model}c.

We note that the energy scale $J=\W$ eliminated in the bond decimation step  is the energy barrier that needs
to be overcome in order
to generate a phase slip which breaks the coherence. The gap to internal fluctuations of the relative phase $\w_J=\sqrt{\W U_i}$ is in general much smaller than $\W$. Fortunately virtual occupation of the Josephson plasmon affects
only a small change on the Josephson coupling of the joined cluster to the
neighboring sites, 
\be J_{i-1}\to J_{i-1}\left(1-{1\over 8}\sqrt{{\tilde U_i}/\Omega}\right),\ee
 which we will find is negligible near the critical point.     
It is important to note here that the above decimation step {\it does
  not} assume long range order; it states that phase
  fluctuations within the newly-formed cluster are harmonic, and
  therefore the cluster can not be broken due to phase-slips. These
  harmonic fluctuations are crucial for the understanding of the
  properties of the superfluid phase, as explained in
  Sec. \ref{SFLutt}.
  
 Before proceeding, we note that in this discussion we have consciously omitted another possible bond decimation step. Suppose that the two sites linked by the strong bond with $J=\Omega$ have large charging energies $U_i>\Omega$, but offset charges $\on_i$ close to $\half$ such that $\D_i<\W$. These sites then have two nearly degenerate charging states, and behave effectively as spin-$1/2$ degrees of freedom rather than rotors. Once connected with a large $J$ of order $\Omega$ they will form a singlet state, i.e., a single boson will be shared equally by them. These sites therefore are not joined to make an effective superfluid cluster, but rather are completely eliminated while generating $J_{eff}$ between the left and right neighbor to the pair. This possibility is treated in Ref. \cite{Altman2010} and is very important in describing the insulating phase. In this review we focus on the superfluid phase and the critical point, where the probability of finding two nearby sites with large $U$ is exceedingly small and can be neglected.  

\subsection{Flow equations and phase diagram}

\subsubsection{Scaling variables}
The iteration scheme outlined above translates  to a flow of distribution
functions for the Josephson couplings, interaction strengths, and offset
charges.
Finding a solution of the flow equations is made easier if we
parametrize the couplings in the Hamiltonain appropriately. Typically,
the best parametrization is in terms of variables that make the
decimation steps become approximate sum rules. For the Josephson
couplings this is achieved by using a logarithmic parametrization: 
\be
\beta_i=\log\W/J_i.
\ee 
The charging energies are best expressed through the capacitance,
which is additive in the cluster formation step. We define:
\be
\zeta_i=\W/U_i.
\ee
Note that $\zeta=2\W C$ with $C$ the capacitance. 
Offset charges are added up as well upon decimation, and therefore
are good variables according to the above criterion. As the
dimensionless flow parameter, we define as before:
\be
\G=\ln \frac{\W_0}{\W},
\ee
where $\W_0$ is of the order of the largest bare energy scale in the Hamiltonian.

\subsubsection{Flow equations}
The distribution functions should reflect which Hamiltonian parameters
are correlated. Clearly $U_i$ and $\on_i$ must be correlated, since
the decimation procedure eliminates sites with large
$\Delta_i=U_i(1-2|\on_i|)$. The elimination of sites with a large
gap, therefore, introduces correlations between charging energies and
offset charges on each site. The Josephson couplings, however, remain
uncorrelated with charging energies or offset charges in nearby
sites. Thus we can parametrize the coupling distribution functions in
terms of two functions: $g_{\G}(\beta)$ the bond log-coupling
distribution, and $F(\zeta,\on)$ the joint distribution function of
the inverse charging energy, and the offset charge. For the latter, it
is useful to write:
\be
F(\zeta,\,\on)=f(\zeta,\on)\Theta(\zeta-1+2|\on|),
\ee
where $\zeta_i=\W/U$ and the Heaviside step function $\Theta$ enforces
the constraint $\W/\Delta>1$ or equivalently $\zeta>1-2|\on|$.

We can now write  the master flow equations for the distributions of coupling constants implied by the decimation rules discussed above:
\bea
\frac{\partial g}{\partial\G}&=&{\partial g\over \partial
\beta}+f_1\,\int d\beta_1 d\beta_2\, g(\beta_1)g(\beta_2)\delta(\beta-\beta_1+\beta_2)+g(\beta)(g_0
-f_1),\label{bgflow1}\\
\frac{\partial f}{\partial \G}&=&\zeta{\partial f\over
\partial \zeta}+g_0 \int d\on_1d\on_2d\zeta_1d\zeta_2 f(\zeta_1,\on_1)f(\zeta_2,\on_2)
\delta(\zeta-\zeta_1+\zeta_2)\delta(\on-\on_1+\on_2)\nn\\&&+f(1+f_1-g_0
).\label{bgflow2}
\eea
Let us  recount the origin of all terms contributing to the derivatives
$dg(\beta)/d\G$ and $df(\zeta,\on)/d\G$. First, a trivial, yet important, effect on the distributions stems from changing the cutoff $\W$. Since $\beta_i=\ln \Omega/J_i$, the change of
$\Omega\rightarrow \Omega-d\Omega$ causes a shift $\beta_i\rightarrow
\beta_i+d\beta$ with $d\beta=d\W/\W=-d\G$. By the same token, this
also shifts $\zeta\rightarrow \zeta + d\zeta$ with $d\zeta=-\zeta
d\W/\W=-\zeta d\G$. These simple shifts are captured through the chain
rule by the first term
in both equations, ${\partial g\over \partial
\beta}$ and $\zeta{\partial f\over \partial \zeta}$.

The next terms are due to the formation of new bonds and clusters. As $\W\rightarrow \W-d\W$, a bond
is decimated if its strength $J$ is within this range, i.e.,
if $\beta<d\G=d\W/\W$. The probability of this for each bond is $g(\beta=0)d\G$.
Out of two clusters with paramters $\zeta_{1,\,2}$ and $\on_{1,\,2}$, a bond decimation produces 
a new cluster with parameters $\zeta=\zeta_1+\zeta_2,\,\on=\on_1+\on_2$
(suppressing the mod in the calculation). This is the essence of the convolution term in Eq. (\ref{bgflow2}). 

\begin{figure}[t]
 \centerline{\resizebox{0.7\linewidth}{!}{\includegraphics{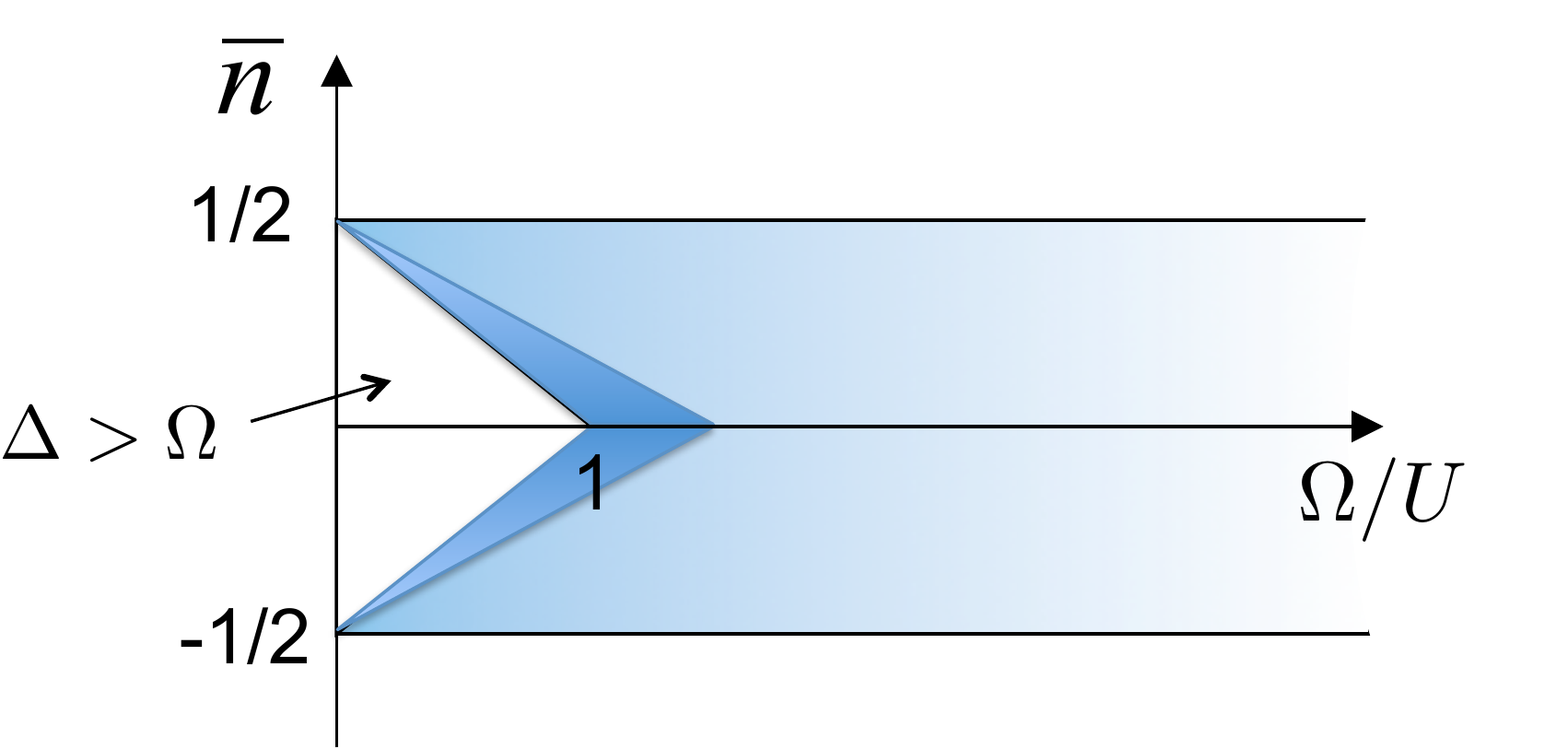}}}
 \caption{\em The region in the parameter space $(\zeta,{\bar n})$ where the charge gap $\D<\Omega$ is the rectangle without the empty triangle. The heavily shaded infinitesimal region will be decimated when $\Omega$ is reduced to $\Omega-d\Omega$. }
 \label{palestinian-flag}
\end{figure}

The convolution term in Eq. (\ref{bgflow1}) describes the formation of a new bond upon site decimation  with the new bond variable  $\b=\b_1+\b_2$ . The probability for such a site decimation event, defined as $f_1 d\G$, is somewhat more complicated than the probability for bond decimation discussed above. This is because a site is decimated when $\W=\D_i=U_i(1-2|\on_i|)$, which occurs at a value of $\zeta$ that depends on $|\bar n|$.  The region in parameter space $(\on,\zeta)$
that would get decimated is shown in Fig. \ref{palestinian-flag} and the corresponding probability is obtained from integrating over it     
\be
f_1=\intt_{-0.5}^{0.5}
d\on\,(1-2|\on|)f(1-2|\on|,\on)
\ee
Note that as the energy
scale $\W$ decreases by $d\W$, the width of the decimated region scales as
$\zeta d\G$, which shrinks to zero on the $\on=\pm1/2$ lines, since $\zeta_i=1-2|\on_i|$ on the decimation curve.

In the vicinity of the transition, the seemingly complicated expression
for $f_1$ simplifies significantly. As more sites are joined, the charge offsets
add up (modulu 1), and therefore rapidly pick values uniformly distributed within $(-\half<\on<\half]$. The distribution
$f(\zeta,\on)$ becomes essentially independent of $\on$, and since it flows
to strong in $\zeta$ it is only weakly dependent on $\zeta$ for values $\zeta<1$. Thus, we can
simply replace $f(1-2\on,\on)\approx f_0$, which implies $f_1\approx \half
f_0$.

The remaining terms in Eqs. (\ref{bgflow1}) and (\ref{bgflow2}) are needed
to maintain the normalization of the distributions. Normalization can be verified by integrating over the
entire range of $\zeta$ and $\on$ in Eq. (\ref{bgflow2}) and $\beta$ in Eq.
(\ref{bgflow1}), and making sure that the derivative of the total probability integral on the
LHS's is zero.

\subsubsection{Scaling solution and reduced flow equations}

Next, we attempt to solve the functional flow equations by employing the
following intuition, which is gained from the spin models analyzed in Sec
\ref{sec:method}. When the decimation steps follow a sum rule for two positive
couplings, the probability distributions for these couplings should consist of exponentials.
This leads to the scaling ansazt:
\bea
 g(\beta)&=&g_0\mathrm e^{-g_0\beta}, \label{ans1}\\
 f(\zeta, \on)&=&{f_0^2\over 1-\mathrm e^{-f_0}}e^{-f_0\zeta}\, \label{ans2}.
\eea
This ansatz describes a family of functions that are parametrized by two
variables, $f_0$ and $g_0$. 
Plugging the scaling ansatz back to the flow equations, one hopes, will generate
a flow equations in terms of $f_0$ and $g_0$ alone, and without any explicit
functional dependence on $\zeta$ or $\beta$. 

\begin{figure}[t]
 \centerline{\resizebox{0.7\linewidth}{!}{\includegraphics{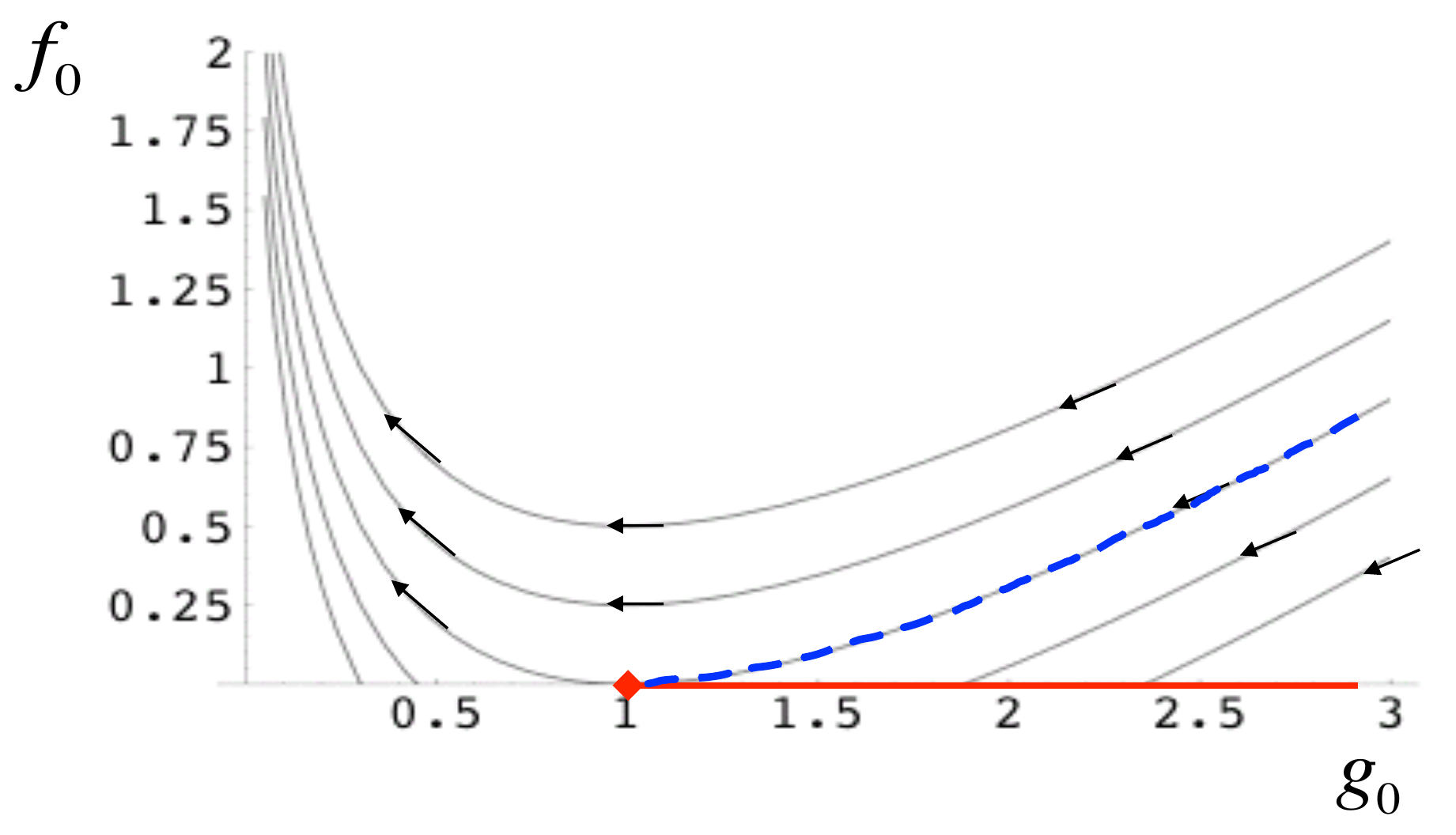}}}
 \caption{{\em RG flow for the reduced variables characterizing the distributions
of Josephson and charging energies. Flows that are below the critical flow line (dashed) terminate on the superfluid fixed line (red).} }
 \label{fig:RGflow}
\end{figure}

Approximately, this is indeed the case. We assume that $f_0\ll 1$, and the
substitution of  the scaling ansatz
to the flow equations yields:
\bea
\frac{df_0}{d\G}&=&f_0 (1-g_0)\label{SFeqa}\\
\frac{dg_0}{d\G}&=&-\frac{1}{2}f_0\, g_0.
\label{SFeq}
\eea
The flow lines in the space $f_0$ versus $g_0$, plotted in Fig. \ref{fig:RGflow}, are given by
\be
f_0=2(g_0-1-\ln g_0)+\e. 
\ee
The different flow lines are parameterized by $\epsilon$, which sets the detuning from the critical manifold at $\e=0$. Flows that lie below the critical manifold, $\e<0$, terminate on the line $g_0>1,\,f_0=0$ marking the superfluid phase.
Values above the critical manifold, $\e>0$, flow to the region $g_0<1$ and where $f_0$
is relevant and flows to larger and larger values. This flow characterizes
the insulating regime of the model. 

We note that the scaling equations (\ref{SFeq}) have the form of the  Kosterlitz-Thouless flow equations,
if we were to rewrite them in terms of $\sqrt{f_0}$ and $g_0$. These scaling variables, however, have a different physical meaning. In particular, the variable that gains a universal value at the transition is the exponent $g_0$ rather than the Luttinger parameter.

Below we elucidate the nature of the superfluid phase.  We then explain the essence of the critical point in light of the anomalous properties of the superfluid leading up to it. This will also serve to clarify why a theory that is perturbative in the disorder strength, such as the replica treatment of Giamarchi Schulz\cite{Giamarchi1987,Giamarchi1988} can fail in this regime.

\subsection{The superfluid phase}

As mentioned above, the line $g_0>1,\,f_0=0$ marks the terminus of the flows
on the superfluid side of the phase diagram. What this fixed line describes,
is the formation of a global superfluid cluster where phase-slips are prohibitively
rare, and do not disturb the superfluidity. Indeed, the RG is dominated by
bond decimations where clusters repetitively coalesce until they connect
the two sides of the chain, and, therefore, the stiffness of this cluster
is finite.

The fact that the interaction parameter $f_0$ flows to zero may at first sight suggest that the superfluid is described by a classical Josephson array with infinite compressibility. To see why this is not the case let us  start with a system of length $L$ and continue the decimation until there is only one site left. This site represents a superfluid cluster extending through the full length $L$ of the chain. The typical value of the charging energy of the whole cluster is $E_c\sim \Omega f_0$, which scales as $\sim 1/L$ in late stages of the RG flow. Hence $f_0$ is irrelevant because it describes the charging energy of clusters that become macroscopic at the end of the flow. So, while the capacitance, which is an extensive quantity, grows as $L$, the compressibility, or capacitance per unit length, approaches a finite value at the terminus of the flow. 

Therefore, a superfluid fixed point, written in terms of the phase variable, is essentially a quantum harmonic theory with random "spring constants" $J_i$ distributed
according to the power-law distribution $P(J)=(g_0/\Omega)\ (J/\Omega)^{g_0-1}$. On the other hand, the masses of the coupled oscillators are related to the compressibility $\kappa$. 

The elementary excitations of the superfluid described by the harmonic theory on the fixed line are harmonic phonons, which are localized by the disorder at all frequencies $\w>0$. For weak disorder the localization length associated with a single phonon wave-function at frequency $\w$ diverges toward zero frequency as $\ell\sim
1/\omega^2$ \cite{Ziman1982}. On the other hand, it was shown in Ref.  \cite{Gurarie2008} that the distributions obtained by SDRG give rise to anomalous localization properties  with:
\be
\ell\sim\l\{\ba{ll} {(\ln^2\omega)}/{\omega} & g_0=1 ~\mbox{critical}\\
{1}/{\omega^{g_0}} & 1<g_0<2 \\
{1}/{\omega^2} &g_0\ge 2 
\ea\rr.
\ee

To see how the wide distributions of $J$ lead to an anomalous superfluid phase, consider the superfluid stiffness  of an effective harmonic chain of length $L$, given by
\be
\rho_s^{-1}={1\over L}\sum_i J_i^{-1},
\label{stiffness}
\ee
According to the SDRG, the random variables $x_i=1/J_i$  are distributed as $p(x)\sim x^{-(1+g_0)}$. For $g_0<2$ the variance of $1/J_i$ diverges, as does the variance of the macroscopic variable $1/\rho_s$. This is the origin of the anomalous behavior in the  superfluid having $1<g_0<2$.
Now we also see why the weak disorder  theory\cite{Giamarchi1987,Giamarchi1988} can fail in the strong disorder weak interaction regime of the superfluid. Consider again the dual lattice action (\ref{GS-latt}) that serves as the starting point for Ref. \cite{Giamarchi1987,Giamarchi1988}. Now, that the disorder in $1/J$ is diverging in magnitude, it cannot be brushed away even if it is perturbatively irrelevant.  

It is important to keep in mind that the power-law distribution with the exponent $g_0$ is not the bare distribution of $J$'s in the array, but rather the fully renormalized distribution at low energy scale. It  is  natural to ask at this point if there is a simple observable  that  can bear witness to this distribution and allow to measure $g_0$. One such quantity is the critical current, which is simply the
Josephson energy of the weakest link in a chain.

Consider a superfluid chain much longer than the coherence length $\xi$, which would diverge at the critical point. Renormalize down to the scale $\xi$, or as we will later show energy scale $\W_\xi=\W_0/\xi$. The dependence of $\xi$ on the detuning from the critical point will be discussed in the next section. At this point we have an effective chain with $N=L/\xi$ links. The Josephson couplings distribute as $P(J)=(g_0 /\W)(J/\W)^{g_0-1} $. The probability density of the weakest link, $J_m$, and of the current $I_c$ (an extreme value statistic), is 
\be
P(I_c)={N \xi \,g_0\over\W_0}{\left( \xi I_c\over\W_0\right)}^{g_0-1}\exp\Big[{-N (\xi I_c/\W_0)^{g_0}}\Big]
\ee
This is a power-law at small values of $I_c$, which is cut of by the exponential. A corollary of this is that the typical value of critical current, defined by the sharp peak of the distribution, vanishes as ${\bar I}_c\sim (\W_0/\xi) (L/\xi)^{-1/g_0}$. That is, it vanishes as a power-law of the system size with power $1/g_0 <1$. 

\subsection{The superfluid-insulator transition \label{SFLutt}}
The critical point which controls the unstable flow toward the insulator sits at $g_0=1$ and $f_0=0$, and marks the end of the superfluid fixed line.
From our discussion above we can gain an intuitive understanding for why the superfluid should break down at $g_0=0$.  If we consider again the  stiffness (\ref{stiffness}) of a classical array (or a harmonic chain), we see that the average $\av{1/\rho_s}$ diverges as $\sim (g_0-1)^{-1}$. Hence, the critical point is where the classical Josephson array loses its stiffness.

It is, however, important to realize that the phase transition is not classical, and the superfluid stiffness at the transition point is not zero. We can think of the transition as being tuned by crossing the critical manifold at some value of $f_0>0$ and $g_0>1$. Quantum fluctuations due to sites with large charging energy drive the downward flow of $g_0$ toward the fixed point, where $g_0=1$. The superfluid stiffness and compressibility, on the other hand, are not universal properties on the fixed point, but must be integrated along the flow.

For example, the stiffness of the final superfluid cluster spanning the chain depends on all the
effective internal bonds $\tilde{J}_i$ that connect the sites making
that cluster and were decimated in the process of its formation:
${\rho_s}^{-1}={L}^{-1}\sum_{i}{\tilde{J}_i}^{-1}$.
Similarly, the compressibility is the sum of capacitances
of all sites making up the superfluid cluster:
$
\kappa={L}^{-1}\sum_i {U_i}^{-1}
$.
In both cases $L$ is the total length
of the chain.
These quantities were computed in Ref. \cite{Altman2010} on the critical manifold showing that they approach a constant value that depends on the initial disorder for the flow.

For the sake of comparison with the weak disorder theory of Giamarchi and Schulz \cite{Giamarchi1987} it is interesting to look at the Luttinger parameter, given by $K=\pi\sqrt{\rho_s\kappa}$.
There, the fully renormalized Luttinger parameter at the transition takes a universal value $K=3/2$. On the other hand, for the strong-disorder transition, the SDRG predicts a value of the Luttinger parameter that depends on where the critical manifold was crossed. That value diverges as the crossing point gets closer to the classical fixed point at $f_0=0$. 

The non universal value of the Luttinger parameter at the transition point is the most controversial prediction of the SDRG in the strong disorder regime\cite{Pollet2013,Pollet2013a}. It is also worth noting that since the Luttinger parameter does not appear as a natural scaling variable in the SDRG, computing its value requires a rather elaborate integration over the entire critical RG flow from high to low energies. Hence it is not surprising that its value on the critical manifold comes out non universal within this theory. But because the Luttinger parameter is not a natural object to compute within the SDRG it is important to verify this prediction in numerical simulations. Such calculations will be discussed in section \ref{sec:numerics}. 

Despite this difference, there is at least a formal similarity between the strong-disorder fixed-point and the Kosterlitz-Thouless flow at weak disorder.  As we already pointed out, the reduced flow equations are formally  identical to the Kosterlitz-Thouless  equations, where the parameters $g_0$ and $\sqrt{f_0}$ of the distribution functions are playing the roles that the Luttinger parameter and phase-slip fugacity play in the Kosterlitz-Thouless flow. 
As a consequence, length and time scales have the same exponential divergence \cite{Altman2004}
\be
\xi\sim \xi_0 e^{a/\sqrt{\epsilon}}~,~\tau\sim \W_0^{-1}e^{b/\sqrt{\epsilon}}
\ee 
as a function of the detuning $\epsilon$ from the critical point, that is a hallmark of the KT transition. 

\subsection{The insulating phases}
The superfluid phase and the superfluid-insulator critical point discussed above are not affected in any important way by the nature of the initial disorder distributions. But type of disorder in the offset charges ${\bar n}_i$ does have a decisive effect on the nature of the insulating phases. Different constraints on the distribution of ${\bar n}_i$ give rise to three distinct insulating phases. The physical properties of these insulators are largely determined by the charging gap distribution of the sites that survive the RG flow deep into the insulating region. The possibilities are as follows:

\begin{itemize}

\item{\em Bose-glass phase --} A generic distribution of offset
charges, $-0.5\le\on<0.5$, results in a Bose-glass phase. This is a compressible and gapless state, which is also  characterized by a diverging 
superfluid susceptibility. The gaplessness and compressibility are due to the presence of a finite density of sites with a charging energy arbitrarily close to zero. The diverging superfluid susceptibility, $\chi$, is a typical property of a compressible phase. The form of the divergence is $\chi_{SF}\sim \ln (\Omega/h_p)$, with $h_p$ the probing proximity field. $\chi$ diverges in the limit of $h_p\rightarrow 0$. Alternatively, for a finite size sample, $\chi_{SF}\sim \ln L$.

\item{\em Mott-glass phase --}  In a commensurate lattice with disorder only in $J$ and $U$, the offset charges vanish identically $\on_i=0$. The system then flows to the
Mott-glass phase; an {\it incompressible} yet gapless state with a finite superfluid susceptibility. Both the compressibility and the gaplessness are a consequence of the universal distribution of $U$, Eq. (\ref{ans2}), which in this case also describes the charging gaps. The distribution $f(U)\sim \frac{1}{U^2}e^{-f_0 \Omega/ U}$ has no support at $U\rightarrow 0$, which implies incompressibility. Specifically: 
\be
\kappa\sim e^{-f_0(\Omega)\frac{\Omega}{\mu}}\rightarrow 0
\ee
where $\mu$ is the probing chemical potential. By looking at the smallest charging energy for a system of size $L$, we find that the gap drops as $\Delta\sim\frac{1}{\ln L}$. The Mott-glass is a classic example of a Griffiths phase: it is incompressible, yet the average temporal auto-correlations of the system will be dominated by the essentially single large cluster of size $\ln L$ that determines the gap. 

\item{\em Random-singlet glass --} If $\on$ is randomly either $\on=0.5$
or $\on=0$, the gapless insulating phase has both a diverging compressibility
and diverging superfluid susceptibility. Only sites with $\on=0.5$ and no charging gap survive late in the flow as two level systems. Thus, the insulator is described by an effective random spin-$1/2$ $xx$ chain, known to be in the random singlet phase \cite{Fisher1994}. Both the compressibility and superfluid susceptibility scale in the same way:
\be
\chi_{SF}\sim\kappa\sim \frac{1}{\mu\ln^3\frac{ \Omega}{\mu}}.
\ee
with $\mu$ being the probing chemical potential {\it or} proximity field. From the random-singlet energy-length scaling, we obtain for a finite size system: $\chi_{SF}\sim\kappa\sim \frac{e^{-c\sqrt{L}}}{L^{3/2}}$.

\end{itemize}

\subsection{Numerical tests of the strong disorder transition}\label{sec:numerics}

The critical point identified using SDRG is characterized by finite randomness. Although the fixed-point analysis is also controlled and justified  by the smallness of interactions in its vicinity, it is not as reliable as the analysis of the random singlet phase or the critical point of the random transverse-field Ising model, where the SDRG flow is to infinite randomness. It is, therefore, desirable to
seek numerical verification of the theoretically predicted universal physics. 
Some of the recent work that addressed this question is summarized by Pollet
in another review published in this volume\cite{Pollet2013a}. In what follows we will survey the numerical results and comment on the debate surrounding their interpretation.

Since we are dealing with a strongly correlated system, a fully quantum calculation, such as quantum Monte Carlo (QMC) or Density Matrix Renormalization Group (DMRG) are needed to extract the critical properties. 
It is important to understand the fundamental difficulties, which complicate these numerical calculations in the strong disorder regime. First, according to the SDRG, rare weak links generated in the course of renormalization play a crucial role in driving the transition at strong disorder. This implies that a very large  ensemble of different realizations of the random chain needs to be analyzed in order to detect the effect of rare events on the physics. Second,  when rare weak links are effective they slow down the convergence of numerical methods such as QMC and DMRG. 

Early QMC studies of the strong disorder regime have not probed directly for the
scaling predictions of the SDRG, but rather checked how BKT scaling works in such systems. For systems with moderate disorder Balabanyan et. al. \cite{Balabanyan2005} found a good fit to BKT scaling with the appropriate value of the Luttinger parameter ($K=2$ for the commensurate system used in that calculation). For stronger disorder, however, the fit did not work as well and strong finite size effects were cited as the reason.  A more recent QMC study \cite{Hrahsheh2012} provided strong, albeit indirect, support for the strong randomness scenario. These
simulations show KT-like scaling of the correlation length at the transition
both for weak and strong disorder. Above a certain disorder strength,
however, the Luttinger parameter as well as the
  susecptibility-length scaling exponent at the
transition were seen to depend on the disorder strength at which the
transition is crossed. In agreement with the strong disorder scenario,
the critical Luttinger parameter exceeded the universal value
predicted by the standard theory\cite{Giamarchi1987}. Analytical
support to the results of Ref. [\onlinecite{Hrahsheh2012}] was given
in  [\onlinecite{Iyer2013}], which also showed, however, that at very
strong disorder the asymptotic value of the Luttinger parameter is
only obtained at very large length scales. 
  
The Luttinger parameter, which was the focus of these studies is, however, not a natural quantity to characterize the strong-disorder critical point. Rather, the relevant scaling properties are encoded in the RG flow, Eqs. (\ref{SFeqa}) and (\ref{SFeq}), of the variables $g_0$  and $f_0$ associated with the distributions of Josephson couplings and charging energies. An apparent difficulty in extracting this scaling behavior from numerical calculations is that the renormalized coupling distributions are not directly observable quantities. This problem was addressed in Ref.  \cite{Pielawa2013}, which developed a finite size scaling theory relying on measurable quantities. A direct  connection was established there between the renormalized distribution of weak links, parameterized by $g_0$, and the measured distribution of superfluid stiffness on an ensemble of finite clusters\cite{Pielawa2013}. Specifically, the tail of the distribution of
the measured value of $1/\rho_s$ in a system of size $L$ was shown to follow a power-law with the same exponent as  the distribution of the variable $x_i\equiv 1/J_i$ at the energy
scale $\Omega(L)$. Thus, the finite-size scaling behavior of the distribution
of $1/\rho_s$, could be compared with finite size scaling formulas
obtained from the RG flow. The QMC results obtained for a model of bosons at integer filling with particle-hole symmetric (off diagonal) disorder were found to scale in the manner predicted by the SDRG to within the numerical error bars. Furthermore, finite size scaling assuming a standard KT transition did not fit the data at all (See Suppllementary material of Ref.  [\onlinecite{Pielawa2013}]).

Another recent study by Pollet et. al. [\onlinecite{Pollet2013}] focused on the anomalous finite-size scaling of the inverse stiffness in the superfluid phase leading up to the critical point. The Monte Carlo simulations carried out in this study also found distributions characterized by broad power-law tails as predicted by the SDRG.
Pollet et al., however, made the interesting observation that the scaling of the median value of $1/\rho$ with $L$ can be fit over a wide regime to the scaling of a classical Josephson array (or a harmonic chain) with a power-law distribution of Josephson couplings $P(J)\sim J^\a$ for small $J$. 

Such a classical JJA, or harmonic chain, have inverse stiffness $\rho_s^{-1}=L^{-1}\sum_i J^{-1}_i$, which leads to the scaling law (up to logarithmic corrections) 
 $\rho_s^{-1}(L)=\rho_s^{-1}(\infty) + a/L^{\a}$ with $\a$ being a length independent exponent. While Ref. [\onlinecite{Pollet2013a}]  did not find noticeable deviations from classical scaling of the median of $1/\rho_s$, the broad power-law distribution of the tail of $1/\rho_s$ is generated in the first place by sites with large charging energy, that is, by a quantum effect. Moreover, the power $\tilde\alpha$ extracted from the tail of the distribution rather than from the scaling of the median of $1/\rho_s$ is seen to flow downward as a function of $L$ due to such site decimations. 
 
Pollet et. al. correctly point out that the tail of the distribution
of $1/\rho_s$ is not important for the thermodynamic stiffness in the
superfluid phase as long as the tail exponent
${\tilde\a}(L\to\infty)>0$. Nevertheless, it is important to note in
this regard that the RG flow of $\tilde\a$ to negative values is one
mechanism that would necessarily destabilize the superfluid. If the
bare disorder is not sufficiently strong compared to the interaction,
then this strong disorder mechanism can be preceded by proliferation
of phase slips through the standard KT mechanism at a universal value
of the Luttinger parameter. Interestingly, for the model, and
parameter regime investigated in Ref. [\onlinecite{Pollet2013a}] the
value of the thermodynamic Luttiger parameter extrapolated in this way
reached the universal value $K_c=3/2$ very close (but on the
superfluid side) of the classical transition point defined by
$\a=0$. 
It is quite striking that, although the critical value of the Luttinger
parameter found in this way is very close to the weak-disorder universal value, the only scaling theory that fits the
data is that predicted by the SDRG. A possible explanation is that
very strong deviations of the Luttinger parameter from the weak
disorder universal value require exceedingly large systems to be
observed according to the scaling analysis in
Ref. [\onlinecite{Iyer2013}] (although values of $K<4$ should still be
reasonably accessible). Regardless, these puzzling results highlight
the need for a more complete understanding of the superfluid-insulator
transition in the intermediate disorder regime, where both
rare weak links and vortex proliferation may play an important role. 
 
\section{Further applications of SDRG to bosonic systems}\label{sec:extensions}

\subsection{Dissipative superconductors}

Systems of Josephson junctions in the presence of dissipation can also be analyzed using SDRG \cite{Hoyos2007}. The analysis considers a large $N$ generalization of Josephson junctions (which would correspond to an $O(2)$ model). In this limit, one can write a quadratic (imaginary-time) action which includes the nearest-neighbor Josphson coupling $J_i$, as well as dissipation $\gamma$, encoded in the frequency-dependent part of the action:
\be 
S=\summ_{\omega}\summ_i \l[-J_i\vec{\phi}_i\cdot  \vec{\phi}_{i+1}+\l(\epsilon_i+\gamma_i|\omega|^{2/z}\rr)\vec{\phi}_i^2\rr].
\ee
The $\epsilon_i$'s are determined self-consistently for each site by requiring that $\langle \vec{\phi}^2\rangle=1$. $z$ is a parameter that determines the nature of the dissipation. The case $z=1$ coincides with the case we have solved in the previous section, with $\gamma$ being the capacitance. $2>z>1$ represents the superohmic regime, while $z>2$ is the superohmic regime. The case of $z=2$ is precisely the Ohmic regime; furthermore, it coincides with the so-called Hertz-Millis theory for fluctuating order-parameters in a dissipative electronic environments. Speicifically, this action describes the onset of supercondcutivity in disordered thin wires \cite{DelMaestro2008}. 

At $z=2$, the ohmic case, one can carry out the same RG steps as outlined in Sec. \ref{rgsteps} for the rotor model. A strong bonds lead to formation of clusters with dissipation $\gamma_{eff}=\gamma_1+\gamma_2$, and decimating a site with a large $\epsilon_2\sim e^{-\gamma_2}$ produces a weak bond between its neighbors, $J_{eff}=2\frac{J_1 J_3}{\epsilon_2}$, where we used sites 1,2 and 3 as an example. This analysis, in the O(N) language, results in a phase transition tuned by $\delta=\overline{\ln J-\gamma}$ between an ordered phase with $J$'s relevant, and a paramagnetic phase with $J$'s scaling to zero. In an electronic wire, this provides a description of the superconducting-metal transition. The critical point is shown to be a random-singlet fixed point, identical in its universal properties to the transverse field Ising model. For the full details of this interesting and surprising result, we refer the reader to Ref. \cite{Hoyos2007}.

\subsection{Two dimensional rotor model}

Applying the SDRG to higher dimensions is always challenging, since the decimations alter the geometry of the system. Nevertheless, several works have succeeded in employing SDRG in higher dimensions for the transverse-field Ising model \cite{FisherMotrunich,RiegerIgloi,MonthusGarel}, the random-hopping model \cite{MotrunichHuseRH}. Recently, a SDRG analysis of the 2d, square lattice, rotor model with no offset charge (as in Eq. (\ref{model}) with $\on_i=0$) was carried out in Ref. \cite{IyerPekkerRefael}. 

Technically, the SDRG method had to be modified in several ways for the 2d rotor model analysis. First, cluster formation led to the addition of Josephson energy in the case of the sites making up the cluster are both connected to another site. Second, because the system may not flow to a strong-disorder fixed point, or one with nearly no interactions (as happens in 1d), the phase fluctuations, i.e., phonons, within clusters were taken into account as suppressing the Josephson couplings emanating from renormalized clusters. Third, in order to make the analysis numerically tractable, only couplings with strength above an absolute cut-off were recorded at each decimation stage. This is particularly important since the connectivity of the system increases dramatically with site decimations. 

The 2d rotor-model SDRG analysis revealed a percolation like superfluid-insulator phase transition. The transition was most apparent by looking at the 
 parameter flow in the space of $\Delta J/\overline{J}$ vs. $\overline{U}/\overline{J}$, with $\overline{J}$ the average Josephson couplings,  $\Delta J$ the standard deviation of $J$, and $\overline{U}$ the average Josephson couplings (see Fig. \ref{2DRGflow}). All quantities were calculated from the largest $2\tilde{N}$ Josephson energies, and charging energies, with $\tilde{N}$ the renormalized size of the system. This is needed to put the analysis on the same footing as the initial model, which only has $2N$ Josephson energies, and $N$ charging energies.  A critical point appears at $\overline{U}/\overline{J}\approx 0.3$ and $\Delta J/\overline{J}\approx 1$. Reassuringly, these values were, by and large, independent of the initial disorder distributions. The transition was established to be  of the percolation type. At the transition, a fractal cluster forms, with a Hausdorff fractal diension of $d_f=1.3\pm0.2$, and a correlation length exponent $\nu=1.09\pm 0.04$. The calculated exponent $\nu$ conforms with the Harris criterion $\nu d>2$. In contrast the standard $XY$ transition relevant for the clean limit is characterized by $\nu\approx 0.663$~\cite{Gottlob1993} and thus violates the Harris criterion.
 
 While the superfluid phase should be conventional, the insulating phase was identified as an incompressible Mott glass, with a gap that falls off as $1/\ln L$ with $L$ the linear size of the system. For complete details we refer the interested readers to Ref. \cite{IyerPekkerRefael}.

\begin{figure}
 \centerline{\resizebox{0.7\linewidth}{!}{\includegraphics{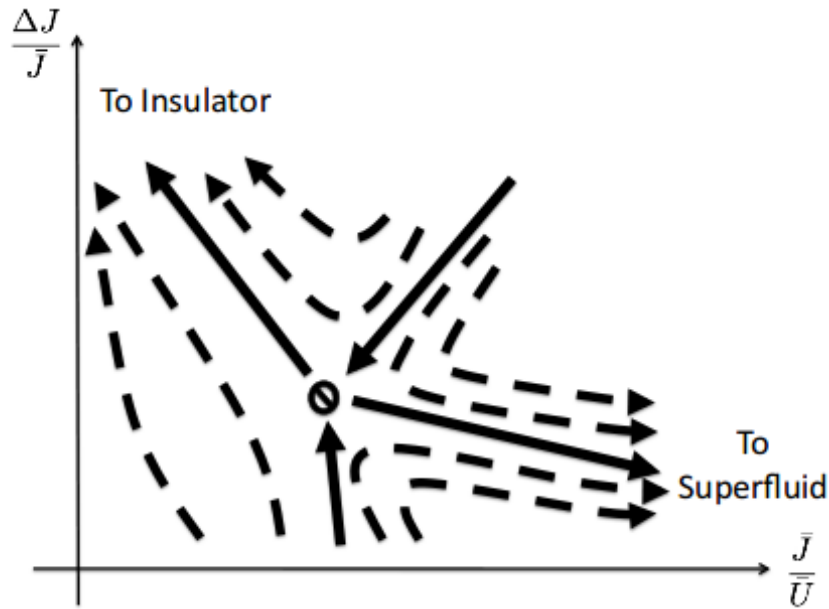}}}
 \caption{The universal flow of the coupling distributions of a 2d rotor model with no offset charges projected onto the disorder vs. the average Josephson to average charging energy ratio. A finite disorder fixed point appears, and can be accessed from both above and below. The flow is obtained using SDRG. It is important to note that for plotting only as many couplings as twice the number of surviving sites were considered.}
\end{figure}
 \label{2DRGflow}

\section{Conclusions}

In this review we have explored two prominent applications of the SDRG method. First, we considered the Heisenberg model, where this method had its first success. Then, we presented the recent application of the method to the random-boson superfluid-insulator transition. The latter problem has been an unresolved problem for a long time, and research complementary to the SDRG analysis is continuing. The question of the ultimate stability of the strong disorder fixed-point to proliferation of phase-slips in the usual Kosterlitz-Thouless mechanism at very long scales remains open. However, numerical simulations indicate that the scaling predictions of the SDRG are, at strong disorder, more relevant to accessible system sizes than those of the standard weak disorder theory. 

Even beyond the problems presented in this review, the application of SDRG to low-dimensional quantum systems continues to be a fruitful research direction. A subjective opinion of the authors, is that the next frontier appears to be the detailed analysis of quantum dynamics \cite{Vosk2013,Pekker2013,Vosk2013a}. The basic principles for all applications of SDRG, however, remain the same as those presented here.

Much of the work reviewed here was carried out together with many collaborators, without whom non of it would have come to pass. We would like to especially acknowledge Daniel S. Fisher, Anatoli Polkovnikov, Yariv Kafri, Victor Gurarie,  Joel Moore, John Chalker, David Pekker, Shankar Iyer, Susanne Pielawa, and Ronen Vosk. In addition, we would like to express our gratitude for the many discussions we had with Thierry Giamarchi, Pierre le Doussal, David Huse, Thomas Vojta, and Steve Girvin, Nikolay Prokof'ev, Boris Svistunov and Lode Pollet. We are grateful to the Packard Foundation, as well as to the NSF, ISF, the Minerva foundation, the Miller Institute of Science at UC Berkeley, and the  Moore Foundation for support through the Caltech IQIM.

\end{document}